\documentclass[12pt,preprint]{aastex}

\shorttitle{Age and Metallicity From Optical and Near-IR Photometry}
\shortauthors{H.-c. Lee et al.}     

\begin{document}

\title{ON THE AGE AND METALLICITY ESTIMATION OF SPIRAL GALAXIES\\
       USING OPTICAL AND NEAR-INFRARED PHOTOMETRY}

\author{Hyun-chul Lee and Guy Worthey}
\affil{Department of Physics and Astronomy, Washington State University, 
    Pullman, WA 99164-2814, USA}

\author{Scott C. Trager}
\affil{Kapteyn Astronomical Institute, University of Groningen, 
    Postbus 800, NL-9700 AV Groningen, The Netherlands}

\and

\author{S. M. Faber}
\affil{UCO/Lick Observatory and Department of Astronomy
    and Astrophysics, University of California Santa Cruz, 
    Santa Cruz, CA 91101, USA}

\begin{abstract}
In integrated-light, some color--color diagrams that use optical and
near-infrared photometry show surprisingly orthogonal grids as age and
metallicity are varied, and they are coming into common usage for
estimating the average age and metallicity of spiral galaxies.  In
this paper we reconstruct these composite grids using simple stellar
population models from several different groups convolved with some 
plausible functional forms of star formation histories at fixed 
metallicity.  We find that the youngest populations 
present ($t<2$ Gyr) dominate the light, and because of their 
presence the age-metallicity degeneracy can be partially broken 
with broad-band colors, unlike older populations. 
The scatter among simple stellar population models by different authors 
is, however, large at ages $t<2$ Gyr.  The dominant uncertainties 
in stellar population models arise from convective core overshoot 
assumptions and the treatment of the thermally 
pulsing asymptotic giant branch phase and helium abundance may play a 
significant role at higher metallicities.  Real spiral galaxies 
are unlikely to have smooth, exponential star formation histories, and 
burstiness will cause a partial reversion to the single-burst case, 
which has even larger model-to-model scatter.  Finally, it is 
emphasized that the current composite stellar population models 
need some implementation of chemical enrichment 
histories for the proper analysis of the observational data.

\end{abstract}

\keywords{galaxies: evolution --- galaxies: photometry --- galaxies: 
spiral --- galaxies: stellar content}

\section{INTRODUCTION}

Much recent works on age estimation of spiral galaxies have used the
surprisingly orthogonal color--color grids that combine optical and
near-infrared colors (e.g., \citealt{BdJ00}; \citealt{mac04}). The
``surprise'' comes from a long list of previous works starting with
\citet{w94} that showed that age effects and metallicity effects in
old stellar populations are very similar on colors, i.e. the
age-metallicity degeneracy problem. Age and metallicity vectors
projected onto color--color diagrams from simple stellar populations
are not orthogonal, but lie nearly parallel to each other so that age
effects and metallicity effects are nigh indistinguishable. The
purpose of this study is to investigate whether the new, seemingly
successful method to break the age-metallicity degeneracy is
convincingly robust or not. The benefits of near-IR photometry
combined with optical photometry have been addressed by many groups
and applied to wide variety of stellar systems (e.g., 
\citealt{dej96}; \citealt{puz02}; \citealt{mac04}; \citealt{lar05};
\citealt{kun05}; \citealt{jam06}). The basic claim is that
age-sensitive main-sequence turnoffs (MSTO) dominate
the visible band while metallicity-sensitive giant branches dominate
the near-IR, and that the photometric behavior of these two
evolutionary regions is sufficiently different to provide diagnostic
power in terms of deriving average ages and metallicities.

It is important to note, however, that the orthogonal grids that
appear in the literature are not simple stellar population (SSP)
models, but composite stellar population (CSP) ones.  The CSP models
that we study here are different from the SSP ones in that CSP models
are made by convolving SSPs with some plausible {\em star formation
histories} (SFHs) at {\em fixed} metallicity and {\em fixed} initial 
mass function.  Obviously, SSP models that consist of a single age
and a single metallicity are not good representations of the
protracted and ongoing star formation history of spiral galaxies.

Before we reconstruct and assess the composite models, however, we
investigate the SSP models in detail because they are the building
blocks of the composite models. We will compare several different
groups' SSP models that have become available lately and look into
their similarities and differences. In particular, we examine the
robustness of stellar population models at ages younger than 5 Gyr
because we perceive that these younger ages dominate the behavior 
of the composite models because of their overwhelming luminosity. 

We then adopt an exponentially declining star formation rate (often
called a $\tau$ model---see Table 1) in order to make composite models.  
We make note of the relationship between SSP and CSP models and what
information flows from the simple to the complex.  Other issues, such 
as different chemical enrichment histories, helium abundance, and
possible variations of initial mass function (IMF) as a function of
metallicity are briefly addressed.

\section{STELLAR POPULATION MODELS}

Many integrated-light models are now available for those interested in
interpreting galaxy spectra and colors.  Since we are starting with the
observational studies of \citet{BdJ00} and \citet{mac04}, we certainly 
must include the stellar evolution upon which the models of 
Bruzual \& Charlot (2003, hereafter BC03) are based because those are 
the models adopted by MacArthur and Bell to build their diagnostic CSP 
models, assuming an exponentially declining SFH at {\em fixed} IMF and 
{\em fixed} metallicity. The rather minor impact of adopting different 
forms of the IMF (e.g., \citealt{sal55}; \citealt{kro02}; \citealt{cha03}) 
is convincingly demonstrated in \citet{mac04}, so we will not explore IMF variation 
in any depth, but will stick with a \citet{sal55} IMF. Besides BC03 
models\footnote{http://www.cida.ve/$\sim$bruzual/bc2003 (based on 
Padova 1994 isochrones)}, we will also examine \citet{rai05} (the 
Teramo group's SPoT\footnote{http://193.204.1.79:21075 (based on the 
earlier version of Teramo isochrones)}), and Maraston (2005, 
hereafter M05)\footnote
{http://www-astro.physics.ox.ac.uk/$\sim$maraston}. 

Both HR diagram behavior and SSP photometric 
models will be examined before we look into CSP models. Of key interest 
are the effects from convective core overshooting and from the 
thermally-pulsing asymptotic giant branch phase at relatively young ages 
(100 Myr to 2 Gyr), so we wish to include a variety of different 
approaches to these aspects of evolution theory.

To do this, we use photometric models that are direct descendents of
an evolutionary population synthesis code that has been developed to
study the stellar populations of star clusters and galaxies
(\citealt{lee02}; \citealt{lee04}) that includes as its evolutionary
basis the $Y^{2}$ isochrones (\citealt{yi01}).  
However, now that we are extending down to 100 Myr, 
the $Y^{2}$ isochrones may not be the best 
choice for this study because there are no post-RGB stellar models yet
for the younger ages.  We include therefore the Teramo groups'
isochrones, BaSTI\footnote{http://www.te.astro.it/BASTI/index.php
(incorporating the last modification of 2006 March 24)} (Pietrinferni
et al.\ 2004; Cordier et al.\ 2007) and the Girardi et al.\ (2000)
isochrones in this study.  We investigate ages from 100 Myr to 13 Gyr 
and metallicities from $Z =0.0003$ to 0.05.  We adopt a 100 Myr limit 
for the most recent epoch of star formation in this study in order to 
accommodate and compare as many different stellar population models as 
possible\footnote{Although BC03 and M05 give their stellar population 
models starting from 1 Myr, BaSTI, for example, present their stellar 
models starting 
from $\sim30$ Myr.  We also note that there are many uncertainties 
such as rotation, mass-loss, magnetic activity to name a few for the 
fine constraints of massive star evolution as well as its 
following supergiant evolutionary 
stage (e.g., Meynet, Eggenberger, \& Maeder 2007).}.

\subsection{HR Diagrams} 

\subsubsection{Effects Due To Convective Core Overshooting} 

We study effects due to convective core overshooting here.  Stars more
massive than $\sim1\,M_{\odot}$, depending on the chemical composition, 
are known to develop a convective core during their hydrogen burning
phases as a result of the strong dependence of CNO cycle energy
generation on temperature. The convective motion in the core could go 
beyond the physical Schwarzschild boundary, carried by momentum.  
In principle, this overshooting phenomenon would make a larger helium 
core, a brighter luminosity, and a longer lifetime because of the 
increased fuel supply.  The chemical composition of the stars determines 
the exact transition mass of convective core overshoot
(e.g., \citealt{dem04}; \citealt{vr06}). Stellar rotation may affect 
it as well (Deupree 1998, 2000; \citealt{mey00}).  For instance,
\citet{gir00} adopt $\sim 1 M_{\odot}$ as the transition mass
independent of chemical composition even though they gradually change
overshooting efficiency between the mass range of $1.0<M<1.5\,M_{\odot}$.  
The comparisons of transition mass treatment among
several different groups are well described in Gallart, Zoccali, \&
Aparicio (2005, supplement section, figure 4).

Figure 1 demonstrates the effects due to convective core overshoot
from the BaSTI.  In this HR diagram stellar isochrones
with and without overshooting effects are depicted at ages 0.1, 1, 5, 
and 13 Gyr at solar metallicity. The tracks labeled `sss' are 
solar-scaled standard isochrones {\em without} convective core overshoot 
and those labeled `sso' are solar-scaled isochrones {\em with}
overshooting effects\footnote{BaSTI employs $1.1\,M_{\odot}$ as the 
transition mass and change the overshooting efficiency between the mass 
range of $1.1<M<1.7\,M_{\odot}$.}.  It is clearly seen 
in Figure 1 that stars with
convective core overshoot (dashed lines) are generally more luminous 
and show less extended blue loop phases at 100 Myr than stars without 
this phenomenon. Moreover, Figure 1 shows that overshooting effects are
more conspicuous at 1 Gyr than at 100 Myr, especially near the MSTO
region in the case of the BaSTI isochrones.  These dissimilarities 
will be converted and displayed as integrated-light color differences 
in Figures 4 and 5.  

It is also interesting to find in Figure 1 that the overshooting 
effects persist even at 5 Gyr for solar metallicity in a minor way. 
However, overshooting effects should disappear altogether at around 
5 Gyr for the metal-poor stars as they become less massive compared 
to those of solar metallicity at the same age. \citet{pie04} note that, 
for ages less than $\sim4$ Gyr, models including overshooting 
provide a ``better match'' to the color-magnitude diagrams of 
star clusters in the Milky Way Galaxy. As stars become less 
massive, however, they develop radiative cores and the issue of 
overshooting effect eventually vanishes, so convective core 
overshooting is only important at ages
$t<5$ Gyr\footnote{In fact, the BaSTI do not provide 
old ($t>10$ Gyr) stellar models with overshooting effects because 
there should not be any differences.}.

\subsubsection{Effects Due To The Treatment Of TP-AGB Stars}

We now investigate the effects of the thermally-pulsing 
asymptotic giant branch (TP-AGB) phase.  The TP-AGB phase is 
the last stage of AGB evolution.  In general, the AGB stars start to appear 
at $\sim 50$ Myr, while RGB stars emerge around 2 Gyr (\citealt{gal05}; 
\citealt{rai05}; \citealt{muc06}). Therefore, after the 
high mass MS stars and supergiants fade away, it is AGB stars 
that dominate the integrated bolometric light until full-fledged 
RGB stars are looming.  The lifetime of AGB stars depends upon their mass 
and composition but is of the order of $10^{7}$ years, 
with about 90\% of the time spent on the early-AGB phase 
and the rest on the TP-AGB.  The TP-AGB phase dominates the light 
in the near-infrared at young ages ($t<2$ Gyr), even though the 
number of stars in this phase is small compared with the number of 
stars in earlier phases.  After $t>2$ Gyr, the RGB tip 
becomes nearly as bright as the AGB tip but is much more 
numerously populated, by nearly a factor of 10 (\citealt{fer04}). 
In Figure 1 (and in Figure 2), the TP-AGB 
phases are depicted with green colors (in the electronic version) 
as described in the BaSTI website\footnote{An additional 250 points 
for the TP-AGB phases on top of the already existing 2000 points were 
recently appended to the Teramo isochrones and are available at the 
BaSTI website (Cordier et al.\ 2007).}.  

Because of the well-known core mass--luminosity relation of AGB stars
(\citealt{pac71}), it is shown in Figure 1 that the luminosity of
TP-AGB stars becomes dimmer as the population becomes older, i.e., the
stellar masses decrease.  Therefore, as Figure 1 graphically
illuminates, the TP-AGB stars are most important at younger ages not
only because they are brighter at high mass but also because RGB
stars do not reach the typical RGB tip luminosity since their
cores are not degenerate. However, once stars go all the way to the
RGB tip and suffer a helium flash, as the population ages (as their
masses become lighter and their cores become degenerate) it is RGB
stars instead of AGB stars that dominate the near-IR flux.  

Recently, several different groups have investigated this interesting 
but complex set of stellar evolutionary phases.  
Vassiliadis \& Wood (1993, hereafter VW93), for instance, predicted 
that the bolometric luminosity of the tip of the AGB should remain 
much brighter than the tip of the RGB. As a matter of fact, 
the brightest stars in some star clusters in the LMC and SMC, 
SWB\footnote{Searle, Wilkinson, \& Bagnuolo (1980)'s 
classification scheme for star clusters in the Magellanic Clouds. 
In general, it is interpreted that star clusters with type I 
are the youngest and that with type VII are the oldest.} IV--VI, 
are carbon-rich AGB stars, and they are generally about one magnitude 
brighter in $K$-band than the tip of the RGB 
(Mould \& Aaronson 1979, 1980; \citealt{fro80}; \citealt{coh81}; 
\citealt{fer04}; \citealt{muc06}). Because of carbon-rich AGB stars, 
the near-IR colors of these star clusters are much redder than 
those of older or younger clusters, whereas the optical colors 
redden more monotonically with cluster age.

In order to incorporate these observational facts, BC03, for example, 
adopted VW93 to extend beyond the early-AGB and took \citet{gro93} 
and \citet{gro95} for the relative lifetimes of carbon-rich TP-AGB stars.  
Recently, \citet{mrg03} have successfully reproduced 
the 2MASS and DENIS observations of the red tail of carbon-AGB stars 
in the LMC theoretically in the ($J-K$, $K$) diagram
using carbon-rich opacities 
instead of the oxygen-rich opacities of M giants.  They indeed 
recognize that carbon stars are about 1 mag brighter in $K$-band 
than oxygen-rich M stars. Historically, it was \citet{bla78} who 
first found that the number ratio of C/M stars sharply decreases 
with increasing metallicity from the SMC, to the LMC, 
to the Galactic Bulge. This is because stars with lower 
metallicities require fewer thermal pulses to change from 
O-rich to C-rich stars (\citealt{liu00}). 

Figures 2 and 3 compare BaSTI and Padova isochrones at 100 Myr and 
at 1 Gyr for 5 different metallicities.  Figure 2 shows 
the solar-scaled, standard (no overshooting) BaSTI, while 
Figure 3 depicts the Padova 2000 isochrones (Girardi et al.\ 2000; 
hereafter Padova00) that adopt overshooting effects as a default. 
There are subtle differences between Figures 2 
and 3 mostly because of the different input physics\footnote{Including 
different choices of $\Delta$Y/$\Delta$Z (BaSTI = 1.4, Padova00 = 2.25).} 
that go into each stellar model. First, 
contrary to the BaSTI-sss isochrones, {\em at 100 Myr} the Padova00 
isochrones become systematically redder with increasing metallicity, 
{\em particularly in the blue loop phases}.
This results in the wide range 
in $V-K$ color as a function of metallicity in the 
simple stellar population models based upon Padova isochrones, such 
as BC03\footnote{BC03 prefer Padova 1994 
isochrones (\citealt{ber94}; hereafter Padova94) to Padova 2000 
ones (\citealt{gir00}; hereafter Padova00) 
because the former better matches galaxy colors with relatively 
{\em low} giant branch temperatures. Otherwise, 
shapes and locations of these two Padova isochrones 
are extensively similar in the HR diagram.}. 
Second, compared to the recent description of full extension along 
the TP-AGB phases of the BaSTI (green lines in Figure 2 
in the electronic version), we deduce 
that the Padova00 isochrones do not have the TP-AGB phase at 100 Myr, 
but at 1 Gyr they go all the way to the TP-AGB luminosity though 
they seem to fail to illuminate carbon stars, as \citet{mrg03} pointed out.
It is also worthwhile to reiterate that Padova00 employs 
convective core overshooting as a default, and that this makes them, 
on the whole, brighter than BaSTI-sss. Moreover, at 1 Gyr, 
the RGBs of Padova00 do not go all the way to the tip luminosity 
because of the overshooting effect, as illustrated in Figure 1.

\subsection{Color--Color Diagrams} 

Having examined the key uncertainties of stellar models 
at young ages in the HR diagram, 
we now compare SSP models in color--color diagram.
The $B - V$ vs. $V - K$ diagrams will be examined first.
This is because there are well established observational datasets 
of star clusters and galaxies in {\em BVK} passbands.
Later we will switch to the familiar $B - R$ vs. $R - K$ plots 
that \citet{BdJ00} and \citet{mac04} presented in their papers.  

Figures 4--7 show several different SSP models 
at 0.1, 1, 5, and 13 Gyr in the $B - V$ vs. $V - K$ diagram. 
At given age, five metallicities are delineated as indicated in 
the bottom right. To guide the eye, solar metallicities are marked 
by filled symbols at given ages. The solid lines with big open squares 
are from BC03.  

In Figures 4 and 5, our SSP computations using the BaSTI 
with and without convective core overshoot are separately plotted at
0.1, 1 and 5 Gyr of age\footnote{Our computations using BaSTI-sss
isochrones are also plotted at 13 Gyr. Following the discussion above, 
isochrones with overshooting will be identical at this age.}. 
Dashed lines with open squares are our models using BaSTI-sss 
and dotted lines with open circles are BaSTI-sso, respectively.  
The only difference between Figures 4 and 5 is in our SSP computations: 
we adopt the BaSTI all the way to the TP-AGB in Figure 4, while 
Figure 5 shows the effect of cutting out the TP-AGB phases 
altogether. In this way, we can see the size of the color shifts arising 
from the TP-AGB contribution (e.g., green lines in Figures 1 and 2). 

The effects due to convective core overshoot are also displayed 
in Figures 4 and 5.  They are dominant, as expected, at young ages only, 
here exemplified at 100 Myr and at 1 Gyr.  In particular, the differences
due to overshooting are more outstanding at 1 Gyr than 100 Myr, as 
expected from the HR diagram from the BaSTI in Figure 1.  Moreover, 
metallicity differences from $Z = 0.0003$ to $Z = 0.004$ at 100 Myr
give rise to significant color differences in Figure 5 
because the MS turoffs and blue loops are considerably hotter 
at $Z = 0.0003$ compared to those at $Z = 0.004$, as illustrated 
in Figure 2.  Furthermore, it should be noted that {\em at 100 Myr} 
the colors from our computations (also from the SPoT in Figure 6 and 
from M05 in Figure 7) 
at solar metallicity are bluer compared to that at $Z = 0.008$
because the blue loop phase of the BaSTI at solar 
metallicity is hotter than the $Z = 0.008$ blue loop\footnote{
The stellar evolutionary phase of the blue loop is notoriously 
difficult to compute and the metallicity seems not the first 
controlling parameter that governs that phase (Santi Cassisi, 
private communication).}.  It is also interesting to find that 
only BC03's 100 Myr models cover a wide range in $V - K$ as a function 
of metallicity: this later will result in the nice orthogonal grids seen 
in the composite models.  By comparing Figures 2 and 3, it can be 
seen that the continual redward march of the blue loop as a function 
of metallicity of the Padova isochrones causes BC03's wide spread in
$V-K$\footnote{Also, bear in mind that BC03's highest metallicity is
Padova94's $Z = 0.05$ compared to Padova00's $Z = 0.03$ that we see in
Figure 3.}.

Comparing Figures 4 and 5, the models lie more closely together 
in Figure 5, especially at 100 Myr, indicating that the
BaSTI's TP-AGB phases contribute substantial 
luminosity in the $K$-band\footnote{BC03 have recently conceded 
that their treatment of TP-AGB phases needs to be revised (see 
Bruzual 2007).}.  There are profound color differences 
caused by convective core overshoot and TP-AGB phenomena, 
especially at 1 Gyr. This can be comprehended by examining the 
HR diagrams in Figures 1 and 2.  At 1 Gyr and solar metallicity, 
for the BaSTI-sss, the TP-AGB phase makes 
$B - V$ only 0.013 mag redder but $V - K$ 0.614 mag redder. 
Without considering those TP-AGB phases, as indicated in Figure 5,
the overshooting effects alone make $B - V$ 0.199 mag bluer and
$V - K$ 0.727 mag bluer at 1 Gyr and solar metallicity. 
They are depicted with vectors in Figure 5. Furthermore,
it is important to note that the BC03 models' highest metallicities go
very red in $V - K$ compared to other SSPs at 5 and 13 Gyr. This is
possibly because BC03's highest metallicity is $Z = 0.05$ with $Y = 0.352$
from the Padova94 assumptions, while the others employ the BaSTI 
at $Z = 0.04$ with $Y = 0.303$. 

In Figure 6, BC03 and SPoT models are compared.  The solid lines with 
open triangles are from Raimondo et al.\ (2005)\footnote{
The SPoT models were developed before BaSTI isochrones have the 
full extension along the TP-AGB phases.  They present 4 different 
TP-AGB cases depending upon mass-loss rates.  We show their B1 models 
with moderate mass-loss rate in this study as posted at the SPoT website.}.  
To guide the eye, solar metallicities are marked with filled symbols.  
By comparing the SPoT models in Figure 6 with our calculations in Figures 4 
and 5, which are similarly based upon the BaSTI, 
it seems that SPoT's TP-AGB treatment (their B1 models 
with moderate mass-loss rate) is in between our 
computations of with and without Cordier et al.\ (2007)'s TP-AGB 
description.  For BC03, it is also most important to note 
that at 100 Myr, the wide range of metallicities merely generates 
small variations, less than 0.2 mag, in $B - V$. This is understandable 
from Figure 3, where the hot upper main-sequence stars counterbalance 
the relatively cool blue loop phase stars to make a conspiratorily 
narrow range in integrated $B - V$ color at 100 Myr.

BC03 models are compared with M05 models in Figure 7.  
Solid lines with big open squares are from BC03 while those with crosses 
are from M05.  To guide the eye, solar metallicities are marked with 
filled and thicker symbols at four ages: 0.1, 1, 5, and 13 Gyr.  
At 100 Myr, M05 models are only given at four metallicities, i.e., 
$Z = 0.001$, 0.01, 0.02, 0.04, so that the bluest color of
M05 at 100 Myr is $Z = 0.001$.  To extend the high metallicity end, M05 
implemented $3.5\, Z_\odot$ metallicity Padova stellar models 
from Salasnich et al.\ (2000) into the fuel consumption theorem-based models.  
Here, we show those M05 SSP models that have a blue horizontal branch 
at the two lowest metallicities.  It is intriguing to note that the most 
serious differences between two models are manifested at 1 Gyr, to a degree
similar to (but more significantly discrepant than) what we have seen
from the model comparisons in Figures 4 and 5.  This is 
because of different description of both the convective core overshoot
and the TP-AGB treatment \citep{mar06}.

\section{COMPARISON OF MODELS WITH OBSERVATIONS}

Having discussed the theoretical uncertainties in stellar models 
and their ramifications on the SSP models, we now present 
a comparison between theoretical models and observational data.
In Figures 8--12, observational data from Large Magellanic Cloud (LMC) 
star clusters are overlaid with BC03, M05, and our computations 
in the [Fe/H] vs. $(B - V)_{0}$ (bottom) and $(V - K)_{0}$ (top) plots. 
LMC star clusters' integrated $(B - V)$ colors are from van den Bergh (1981).  
Their reddening and $(V - K)_{0}$ and SWB types are from Table 4 
of Persson et al.\ (1983).  The ages and metallicities of these 
star clusters are adopted from Table 2 of Pessev et al.\ (2006).  
We note, however, that recent measurements of higher metallicity 
LMC clusters ([Fe/H] $>-1.0$ dex) by Grocholski et al.\ (2006) show 
a very tight metallicity distribution (mean $\mathrm{[Fe/H]} = -0.48$, 
$\sigma=0.09$).  Among SWB I--III types, clusters younger than 100 Myr 
are depicted with open triangles.  

According to \citet{fro90}, the SWB type V star clusters in the LMC have 
bright AGBs with carbon stars (see also Mucciarelli et al.\ 2006 for recent 
near-IR CMDs). Indeed, the $V - K$ colors of some of these LMC SWB type V 
star clusters (solar symbols in Figures 8--12) are redder than the average 
trend.  The observational scatter in $V - K$ color of LMC SWB 
types I--III star clusters should be understood by some other mechanism, 
such as stochastic effects \citep{sf97}, but certainly not by 
metallicity (see also \citealt{gou06}).  One of the old SWB VII clusters, 
NGC 1841, is also indicated in the plots 
because of its unusually red $(B - V)_{0}$ color.  
Its $(B - V)$ is 0.90 according to Table 3 of van den Bergh (1981) and 
its $E(B - V)$ is given as 0.07 in Table 4 of Persson et al.\ (1983).  
However, its peculiarity can be removed if we adopt 
0.80 for its $(B - V)$ from Table 1 of Bica et al.\ (1996) and 
0.18 for its $E(B - V)$ from Walker (1990).  

Now we take a close look how different SSP models are compared with 
these LMC star clusters with wide range of age and metallicity.  
First of all, BC03 models are overlaid with LMC star clusters in Figure 8.  
Those four ages (100 Myr, 1, 5, 13 Gyr) that are shown at the previous 
color--color plots are delineated with lines with crosses, triangles, 
squares, and circles, respectively.  To add detail, we also display model 
predictions at finer age spacings.  At solar metallicity, eight 100 Myr 
increments between 100 Myr and 1 Gyr, and 1.5, 2, 3, and 4 Gyr between 
1 Gyr and 5 Gyr are connected with dots.  {\em Both} the $B - V$ and $V - K$ 
colors of BC03 models become {\em monotonically} redder with increasing 
age at given metallicity.  Kerber, Santiago, \& Brocato (2007) recently 
determined ages for four of the intermediate-age star clusters using deep 
HST color-magnitude diagrams (CMDs).  Using Padova models, ages of NGC 1856, 
NGC 1831, NGC 2173, and NGC 2121 were estimated as 300 Myr, 700 Myr, 1.6 Gyr, 
and 2.9 Gyr, respectively, and they are compared with models in Figures 8--12.  
It is noted that contrary to $B - V$ colors, $V - K$ colors show some evident 
mismatch between intermediate-age LMC star clusters and model 
predictions (see footnote 15).  

Secondly, in Figure 9, M05 models are compared with LMC star clusters.  
Just like the previous figure, four ages (100 Myr, 1, 5, 13 Gyr) that are shown 
at the Figure 7 are delineated with thicker lines with crosses, triangles, 
squares, and circles, respectively.  As we mentioned earlier, for a close 
examination, we now show every 100 Myr increment between 100 Myr and 1 Gyr, 
and 1.5, 2, 3, and 4 Gyr between 1 Gyr and 5 Gyr from her models.  
We have used crosses, triangles, squares, and circles repeatedly 
for the description of increasing ages.  While her $B - V$ colors become 
monotonically redder with increasing age at given metallicity, her $V - K$ 
colors show a sudden leap at 300 Myr because of the onset of the TP-AGB.  
Note that M05 tuned her fuel consumption theorem parameters using the 
older age estimation of intermediate-age LMC star clusters.  
The exact epoch of onset will be satisfactorily constrained in the near 
future with more number of detailed deep CMDs of nearby young and 
intermediate-age star clusters along with the help of improved stellar models 
(e.g., Kerber et al.\ 2007; Mucciarelli et al.\ 2007).

Finally, our calculations with BaSTI are presented in Figures 10--12.  
Figure 10 shows our calculations employing BaSTI-sss isochrones 
with the exclusion of TP-AGB phases, while Figure 11 
illustrates those with TP-AGB phases.  It is seen that 
the $V - K$ colors become significantly redder at younger 
ages because of the TP-AGB phases, whereas they do not affect $B - V$ 
colors much.  The apparent mismatch in $V - K$ colors between our 
calculations and intermediate-age LMC star clusters lingers 
even with TP-AGB phases.  In Figure 12, we add convective core 
overshooting effects along with a TP-AGB using BaSTI-sso isochrones.  
It is seen that both $B - V$ and $V - K$ colors become bluer compared to 
Figure 11 at younger ages because of the convective core overshooting 
effects.  Although this case should be the most realistic one, 
it does not ameliorate the mismatch between observations and models 
at $V - K$ colors.  It is believed though that all these discrepancies 
will be narrowed down with the improvement of stellar 
models (e.g., Marigo 2007).  

Having studied the model differences and their matches 
against the observations in $BVK$ photometry, 
in Figure 13, we switch from $B - V$ vs. $V - K$ to 
$B - R$ vs. $R - K$, as \citet{BdJ00} and \citet{mac04} presented 
their spiral galaxy data in these colors.  Figure 13 is similar to Figures 
6 and 7, but here BC03, M05, and SPoT SSP models at 0.1, 1, 5, and 13 Gyr 
are compared in the $B - R$ vs. $R - K$ diagram.  This is the basis 
of Figure 14, the composite model grids when the Figure 13 SSP models 
are convolved with exponentially declining SFHs.  Additionally in 
Figure 13, BC03's solar metallicity {\em composite} models from Figure 14 
are shown as big filled squares connected by a thick line 
(indicating the 7 average ages, $\langle A \rangle$, in Figure 14) to compare 
SSP and CSP models. According to Fig.\ 11 of \citet{mac04}, spiral 
galaxies with {\em high} rotational velocity mostly 
swarm around this thick line.

\section{RESULTS \& DISCUSSION}

Composite stellar population models from three different groups 
(BC03, M05, SPoT) are shown in Figure 14.  BC03 (black) and M05 (red) 
are compared in the top panel and BC03 and SPoT (blue) are contrasted 
in the bottom panel.  We have generated each model 
grid using {\em identical} star formation rates (SFR).  
The simple exponential SFR that we use in this study is
$\Psi_{\mathrm{exp}}(t)$,
\begin{equation}
\Psi_{\mathrm{exp}}(t) = \frac{c_{\mathrm{exp}}}{\tau} e^{-t/\tau},
\label{eq:exp}
\end{equation}
where
\begin{equation}
c_{\mathrm{exp}} = \frac{1}{1 - e^{-A/\tau}}
\end{equation}
and $A = 13$ Gyr.  
The average age by which they are labeled is a {\em mass-weighted} 
average age and at a given $\tau$, computed as:  
\begin{equation}
\langle A \rangle = 
             A - \tau \frac{1-e^{-A/\tau}(1+A/\tau)}{1-e^{-A/\tau}}.
\label{eq:avgageexp}
\end{equation}
For example, a constant star formation rate over 13 Gyr yields an average 
age of 6.5 Gyr. If weights by luminosity rather than mass, one arrives at 
much younger ages. For comparison, we list $V$- and $K$-band 
luminosity-weighted ages at solar metallicity for BC03 models along with 
the $\tau$ and mass-weighted ages in Table 1.  Our mass-weighted age is 
actually the $\langle A\rangle_{cutoff}$ that appears in the third column 
of Table 1.  This is because we have considered the most recent star 
formation to be truncated at 100 Myr in this study\footnote{
Besides the reason that we noted in Section 2, it is seen in Table 1 
that only the $\langle A\rangle=5.1$ Gyr case is affected by the 100 Myr 
minimum age, and the effect is negligible.  
Although there is ongoing star formation ($t<100$ Myr) in spiral 
galaxies, that mass-fraction is not usually a major component 
(e.g., Hammer et al.\ 2005).}.  
In Table 1, the luminosity-weighted ages are indeed 
much younger than the mass-weighted ages because young stellar components 
dominate the light in the composite populations, as can be seen from 
Figure 13 when SSP ages are compared to the CSP ages. 

The models illustrated in Figure 14 are clearly different from one another, 
although they seem to agree to a certain degree at solar metallicity 
(delineated with thick lines to guide the eye). 
The differences arise from the original dissimilarities of simple 
stellar populations. In fact, the averaging process for the $\tau$ 
model CSPs smooths over many of the angularities present 
in the SSP grids to yield something more regular. 
It is also worthwhile to note that these composite
grids are built {\em at fixed metallicity}, without any chemical evolution. 
The behavior of the CSP models echoes that of the SSP ones, and especially 
that of the younger components because younger subpopulations are 
brighter and tend to dominate the light.  In the case of M05 CSP models, the 
iso-metallicity lines of $Z = 0.01$ and 0.02 even cross over, echoing 
their corresponding SSP models at 100 Myr and 
at 1 Gyr. The SPoT's CSP grids are much more compact in $R - K$ colors 
compared to BC03 both at the metal-poor and at the metal-rich 
side, also echoing their original SSP models. 

We have used the same symbols for similar metallicities in Figure 14 
in order to make the comparisons easy.  The highest metallicities of three models 
are almost the same: the highest metallicity in BC03 is $Z = 0.05$ with 
$Y = 0.352$ while the two other are $Z = 0.04$ with $Y = 0.303$ (SPoT), and 
0.340 (M05). The thick lines in Figure 14 are solar iso-metallicity 
lines for each CSP models. We hope that the fact that different models 
become similar at this rather critical metallicity is a robust result 
and not a coincidence.  Reading from Figure 14, a galaxy with 
colors $B - R = 0.97$ and $R - K = 2.37$ would, in the M05 $\tau$ models, 
have an age of 5.1 Gyr and $Z = 0.04$, or $\mathrm{[Fe/H]} = +0.3$ 
where [$\alpha$/Fe] = 0.  
The same galaxy would have $(\mathrm{Age}, Z, \mathrm{[Fe/H]}) = 
( 7.0, 0.037, 0.27 )$ according to BC03, or $(6.7, 0.052, 0.41)$ according 
to SPoT. Taking another galaxy as an example, with $B - R = 1.61$ and $R - K 
= 2.39$, M05 would indicate $( \mathrm{Age}, Z, \mathrm{[Fe/H]} ) = 
( 13, 0.02, 0.00 )$, while BC03 indicates $(15, 0.008, -0.40)$ and 
SpoT indicates $(16, 0.012, -0.22)$.  The uncertainties appear to be 
in line with those tabulated in Charlot, Worthey, \& Bressan (1996); 
about 30\% in age.  The scatter in abundance may be a bit larger than 
the 0.1 dex predicted by Charlot et al.\ (1996).  
In terms of relative, $\Delta Z / \Delta \mathrm{Age}$ measurements, 
such as would be applied to spatial gradient data, 
Figure 14 yields similar percentage scatter.

Notwithstanding the incorporation of star formation history using the
exponentially declining $\tau$ models, Figure 14 is still far from 
a realistic description of stellar populations of spiral galaxies. 
This is because even models like in Figure 14 do {\em not} yet include any 
form of chemical evolution because they are built at {\em fixed} 
metallicity.  In Figure 15, we experiment the same exponential SFR 
that we have adopted in Figure 14, but with one very simple and 
useful case of {\em chemical enhancement} where metallicity is varied 
along with age.  For this experiment, $Z = 0.0001$ 
is assigned for the stars that formed at 13 Gyr ago, $Z = 0.0004$ for the 
stars formed at 11 and 12 Gyr ago, $Z = 0.004$ for the stars formed at 9 
and 10 Gyr ago, $Z = 0.008$ for the stars formed at 7 and 8 Gyr ago, 
$Z = 0.02$ for the stars formed at 2, 3, 4, 5 and 6 Gyr ago, and 
$Z = 0.05$ is given for the stars that formed at 0.1 and 1 Gyr. 
These models are marked by big circles in Figure 15, one for each 
{\em mass-weighted} average age, $\langle A\rangle$. 
In this case with strong age-metallicity relation, the CSP models 
with older {\em average} ages become considerably bluer 
in $R - K$ because of their (assumed) low metallicity 
as most star formation happens at very early cosmic time. Models 
with younger average ages, on the other hand, become progressively 
redder because recent star formation has high metallicity. 

These models are clearly far from the observations of spiral 
galaxies (roughly located near the thick line in Figure 15 
according to Fig.\ 11 of MacArthur et al. 2004) suggesting that this toy model 
with a strong age-metallicity relation is incorrect.  However, 
the toy model serves to illustrate that a general trend caused 
by {\em any} chemical evolution scheme with an age-metallicity relation 
will cause a {\em tilt} in the model grids.  The toy model is an extreme 
case, but almost any chemical evolution scheme will share the 
characteristic that star formation starts at low metallicity at early times 
and works toward high metallicity at late times. This will, in every case, 
induce a bias in the CSP color-color plots.  Clearly, future work should 
include chemical evolution if coherent age results are the goal.

To summarize our results, the youngest populations in {\em composite} 
stellar population models dominate the light and act to defeat the 
otherwise universal age-metallicity degeneracy in color--color diagrams 
that employ optical and near-infrared photometry.  But scatter from model 
to model is large at simple stellar population ages $t<2$ Gyr.  The dominant 
uncertainties in stellar population models arise from the convective 
core overshoot and the TP-AGB phase, but helium abundance may play a 
significant role at higher metallicities.  The various models show notable 
differences that result in considerably dissimilar composite grids in 
integrated light.  These results suggest that the interpretation of 
rest-frame near-IR photometry is severely hampered by model 
uncertainties and therefore that the determination of ages 
and metallicities of spiral galaxies from optical and near-IR 
photometry may be harder than previously thought.  
Another source of cosmic variation could be introduced by the character 
of the star formation; whether it is smoothly varying like the models or 
characterized by bursts.  Burstiness will cause a partial reversion 
to the SSP case, which has larger model-to-model scatter than the CSP 
models.  Furthermore, it is emphasized that the current composite 
stellar population models do need some implementation 
of chemical enrichment histories for full interpretation of 
the observational data.

\acknowledgments

It is a great pleasure to thank Lauren MacArthur, Santi Cassisi, 
Gustavo Bruzual, Claudia Maraston, Mustapha Mouhcine, 
Gabriella Raimondo, Alessandro Bressan, Roelof de Jong, Aaron Dotter 
and John Blakeslee for many helpful discussions.  We also thank the 
anonymous referee for her/his thoughtful comments and insightful 
suggestions that improved this paper greatly.  Support for this work 
was provided by the NSF through grant AST-0307487, the New Standard 
Stellar Population Models (NSSPM) project.

\clearpage

\begin{deluxetable}{ccccc}
\tabletypesize{\scriptsize}
\tablecaption{Comparison of average ages at solar metallicity for BC03 models}
\tablewidth{13cm}
\tablehead{
\colhead{$\tau$\tablenotemark{a}} & \colhead{$\langle A\rangle$\tablenotemark{b}} & \colhead{$\langle A\rangle_{cutoff}$\tablenotemark{c}} & \colhead{$\langle A\rangle_{V}$\tablenotemark{d}} & \colhead{$\langle A\rangle_{K}$\tablenotemark{e}} \\
}
\startdata
 $-10$ & 5.1 & 5.2 & 1.8 & 3.0 \\
 500 & 6.5 & 6.5 & 2.6 & 4.2 \\
 13 & 7.6 & 7.6 & 3.5 & 5.4 \\
 6.5 & 8.5 & 8.5 & 4.8 & 6.7 \\
 4 & 9.5 & 9.5 & 6.7 & 8.2 \\
 2.6 & 10.5 & 10.5 & 9.1 & 9.8 \\
 0.1 & 12.9 & 12.9 & 12.9 & 12.9 \\
\enddata
\tablenotetext{a}{It may be useful to refer to the top portion of figure 4 
of MacArthur et al.\ (2004) for the shape of star formation rate over ages 
with given $\tau$.  A couple of cases of bursty SFHs are also demonstrated 
in figures 6 and 7 of MacArthur et al.\ (2004).}
\tablenotetext{b}{Mass-weighted average age in Gyr.}
\tablenotetext{c}{Mass-weighted average age in Gyr with 100 Myr cutoff.}
\tablenotetext{d}{$V$ Luminosity-weighted average age in Gyr.}
\tablenotetext{e}{$K$ Luminosity-weighted average age in Gyr.}
\end{deluxetable}

\clearpage

\begin{figure}
\epsscale{1.}
\plotone{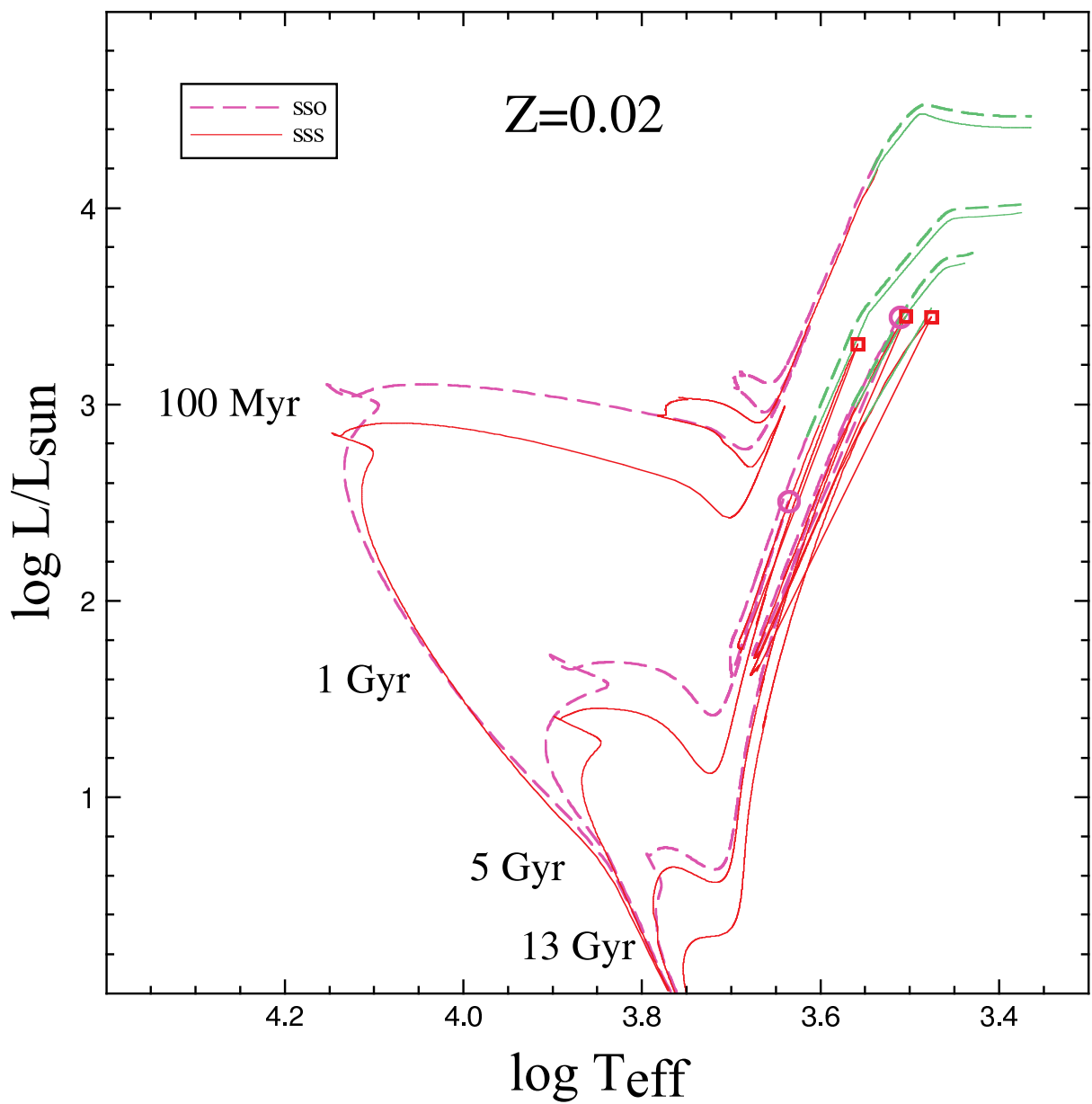}
\caption{Teramo isochrones (BaSTI) are illustrated  at 0.1, 1, 5, and 13 Gyr 
at solar metallicity.  Those labeled `sso' are solar-scaled isochrones 
{\em with} convective core overshoot effects (dashed lines), while 
the `sss' are solar-scaled isochrones {\em without} overshooting 
effects (solid lines).  The RGB tips are marked by 
squares (without overshooting) and circles (with overshooting) at 
1, 5, and 13 Gyr.  In the electronic journal, one can see the 
full extension along the TP-AGB phases 
of the BaSTI in green.}
\end{figure}

\begin{figure}
\epsscale{1.}
\plotone{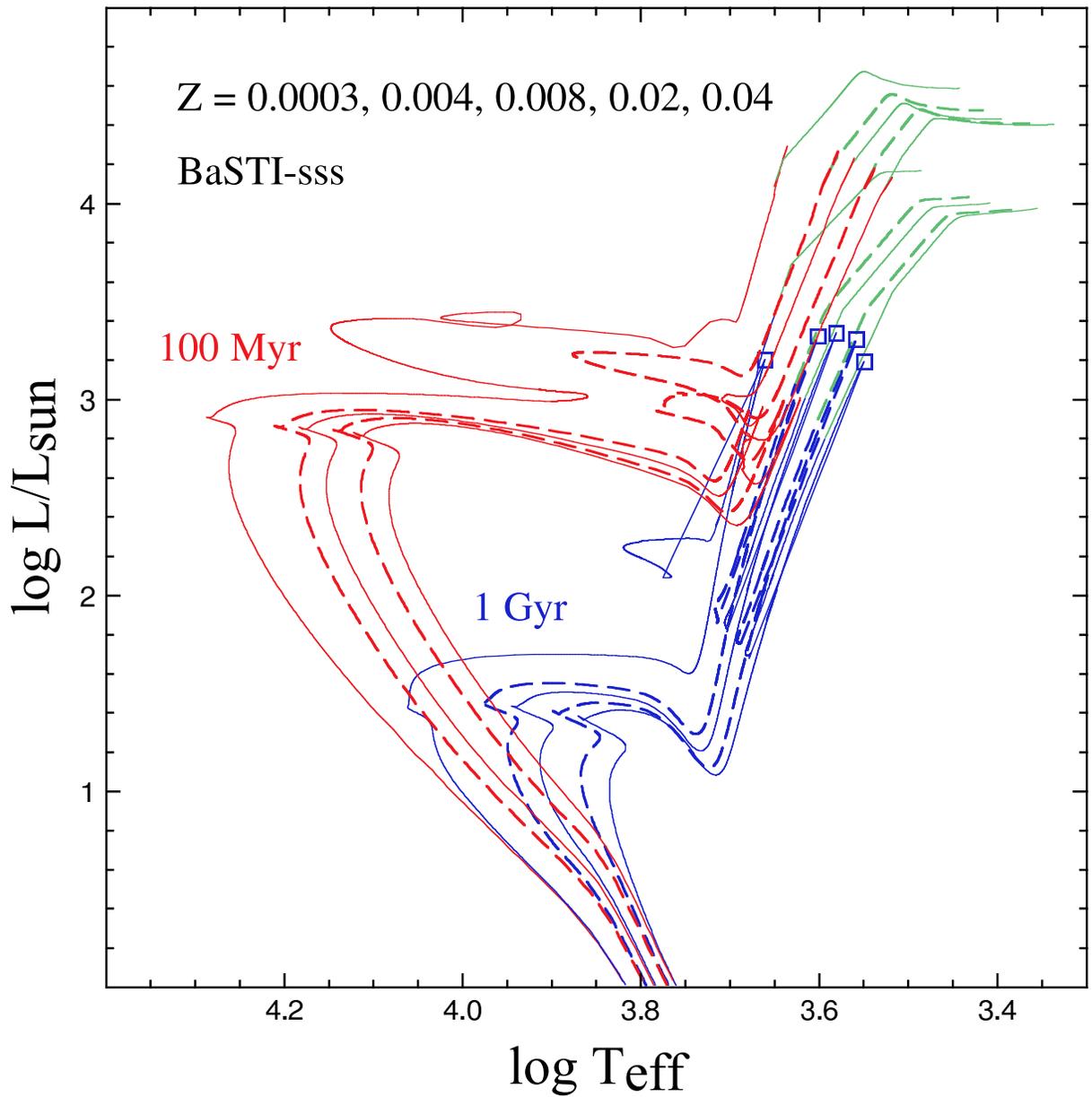}
\caption{BaSTI-sss (solar-scaled standard; no overshooting) 
isochrones are compared at 100 Myr and at 1 Gyr for the 5 metallicities 
that we illuminate in this study (Figures 4, 5, 10, and 11).  
To guide the eye, $Z = 0.004$ and $Z = 0.02$ are 
drawn with dashed lines.  Also, the RGB tips are marked by squares at 
1 Gyr. Note that the blue loop phase at solar metallicity locates 
at the hotter side compared to that of $Z = 0.008$ at 100 Myr in this 
isochrones sets.  In the electronic journal, one can see the 
full extension along the TP-AGB phases 
of the BaSTI in green.}
\end{figure}

\begin{figure}
\epsscale{1.}
\plotone{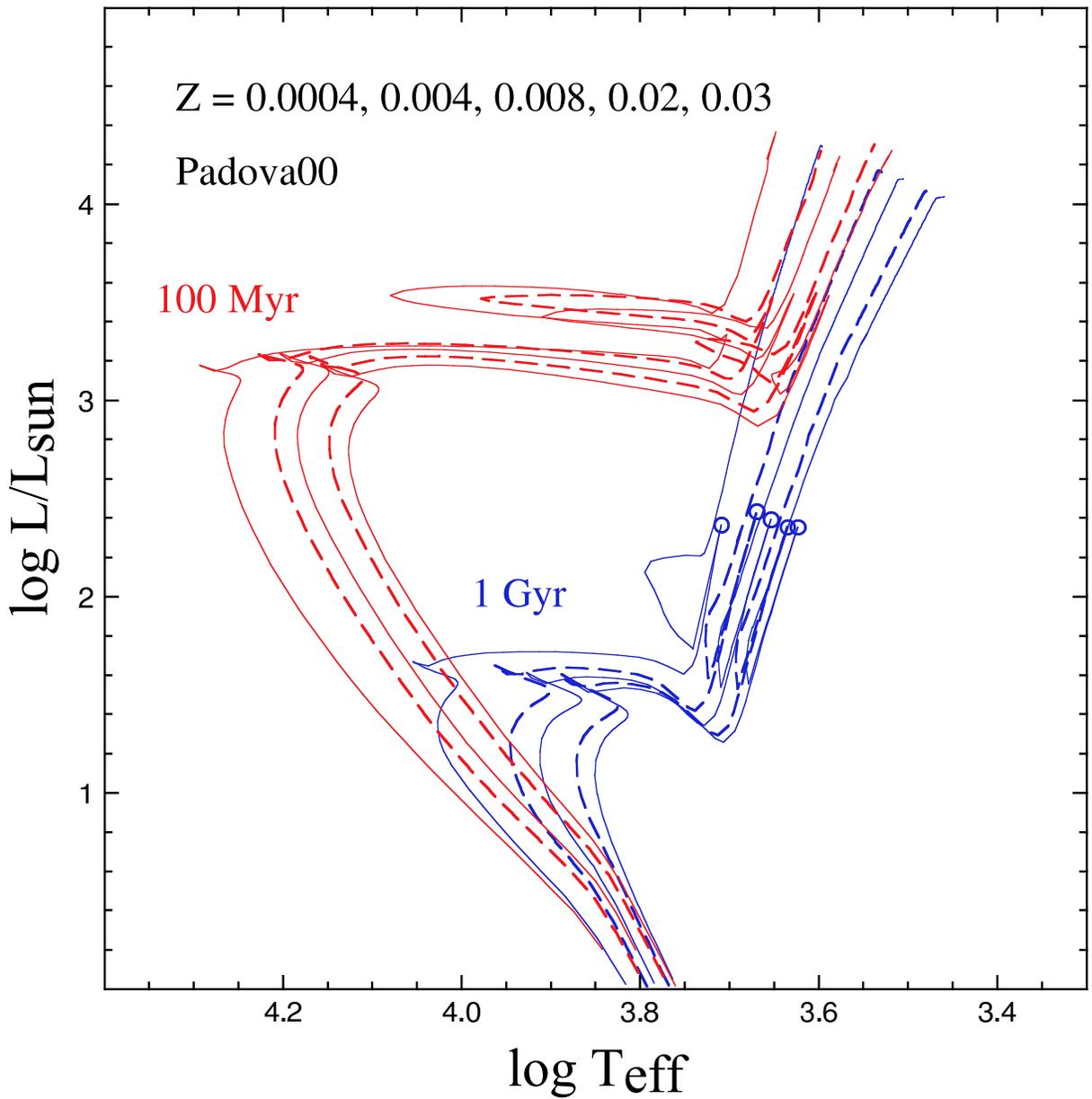}
\caption{Similar to Figure 2, but Padova00 isochrones 
are displayed at 100 Myr and at 1 Gyr for the 5 metallicities. 
To guide the eye, $Z = 0.004$ and $Z = 0.02$ are drawn with dashed 
lines.  The RGB tips are marked by circles where they 
exist (at 1 Gyr), but they do not go all the way 
to $\log L \approx 3.2$ due to convective core overshoot effects.  Note 
also that this isochrone sets are relatively brighter than the ones in 
Figure 2 because of overshooting.  Moreover, the blue loop phases 
at 100 Myr are systematically shifting to the right with 
increasing metallicity compared to the ones in Figure 2.}
\end{figure}

\begin{figure}
\epsscale{1.}
\plotone{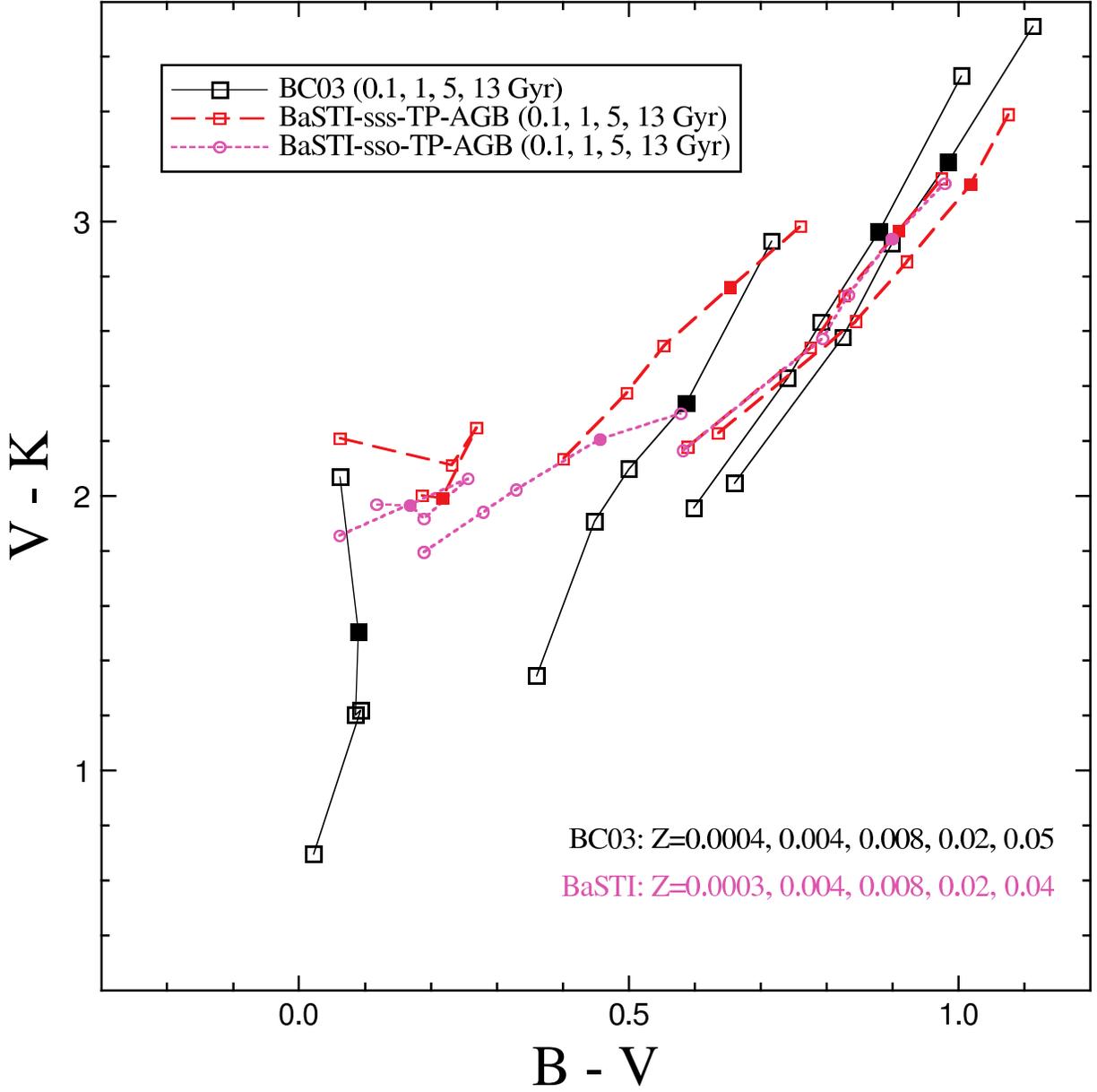}
\caption{($B - V$) vs.\ ($V - K$) plot of simple stellar population 
models at 0.1, 1, 5, and 13 Gyr.  Lines connect different metallicities 
at given ages. Solid lines with big open squares are from BC03 
(also shown in Figures 5, 6, and 7).  Dashed lines 
with small open squares are our computations using BaSTI-sss 
isochrones {\em including} TP-AGB phases and dotted lines with open 
circles are those using BaSTI-sso isochrones.  At 13 Gyr, only the BaSTI-sss 
case is shown because there is no overshooting effect at 13 Gyr.  
To guide the eye, solar metallicity models are marked by filled symbols.}
\end{figure}

\begin{figure}
\epsscale{1.}
\plotone{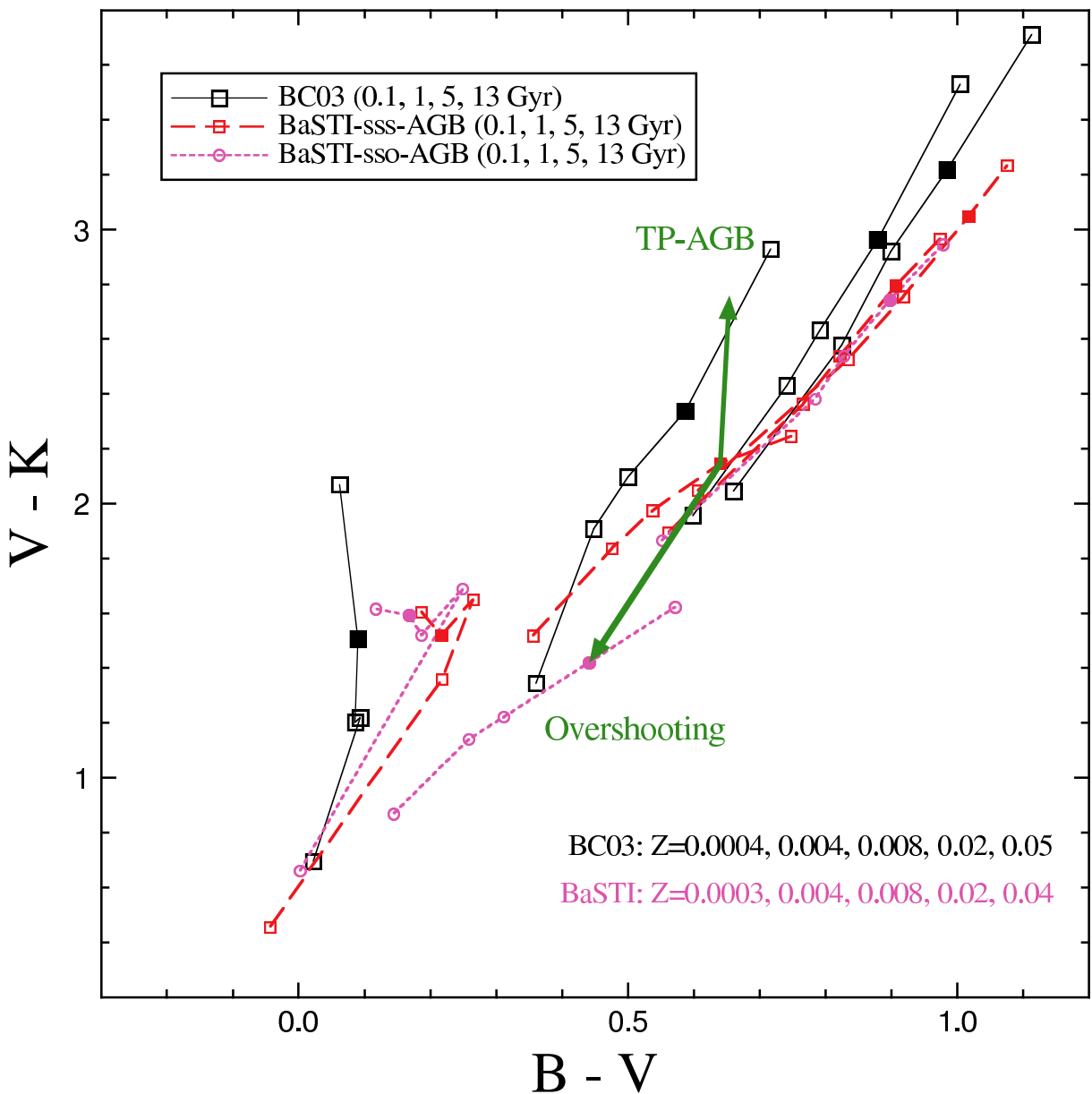}
\caption{Same as Figure 4, but our computations using the BaSTI-sss 
and BaSTI-sso isochrones {\em without} TP-AGB phases are shown as 
dashed and dotted lines, respectively.  At 1 Gyr and solar metallicity, 
for the BaSTI-sss isochrones, the TP-AGB and the overshooting effects 
are depicted with vectors.}
\end{figure}

\begin{figure}
\epsscale{1.}
\plotone{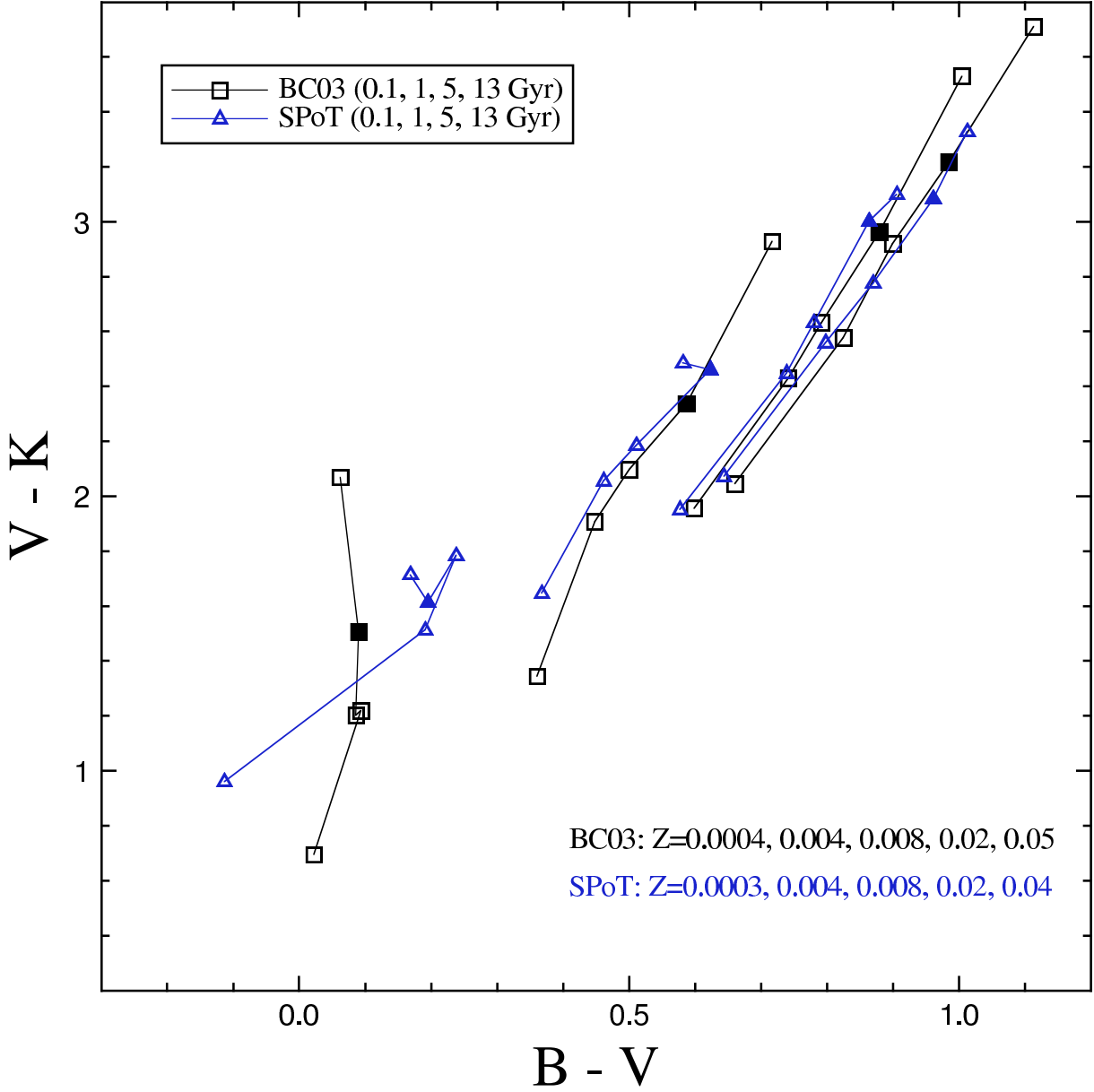}
\caption{Similar to Figure 4, but the SPoT models 
(solid lines with open triangles) are compared with BC03.  
To guide the eye, solar metallicity models are marked by filled symbols.}
\end{figure}

\begin{figure}
\epsscale{1.}
\plotone{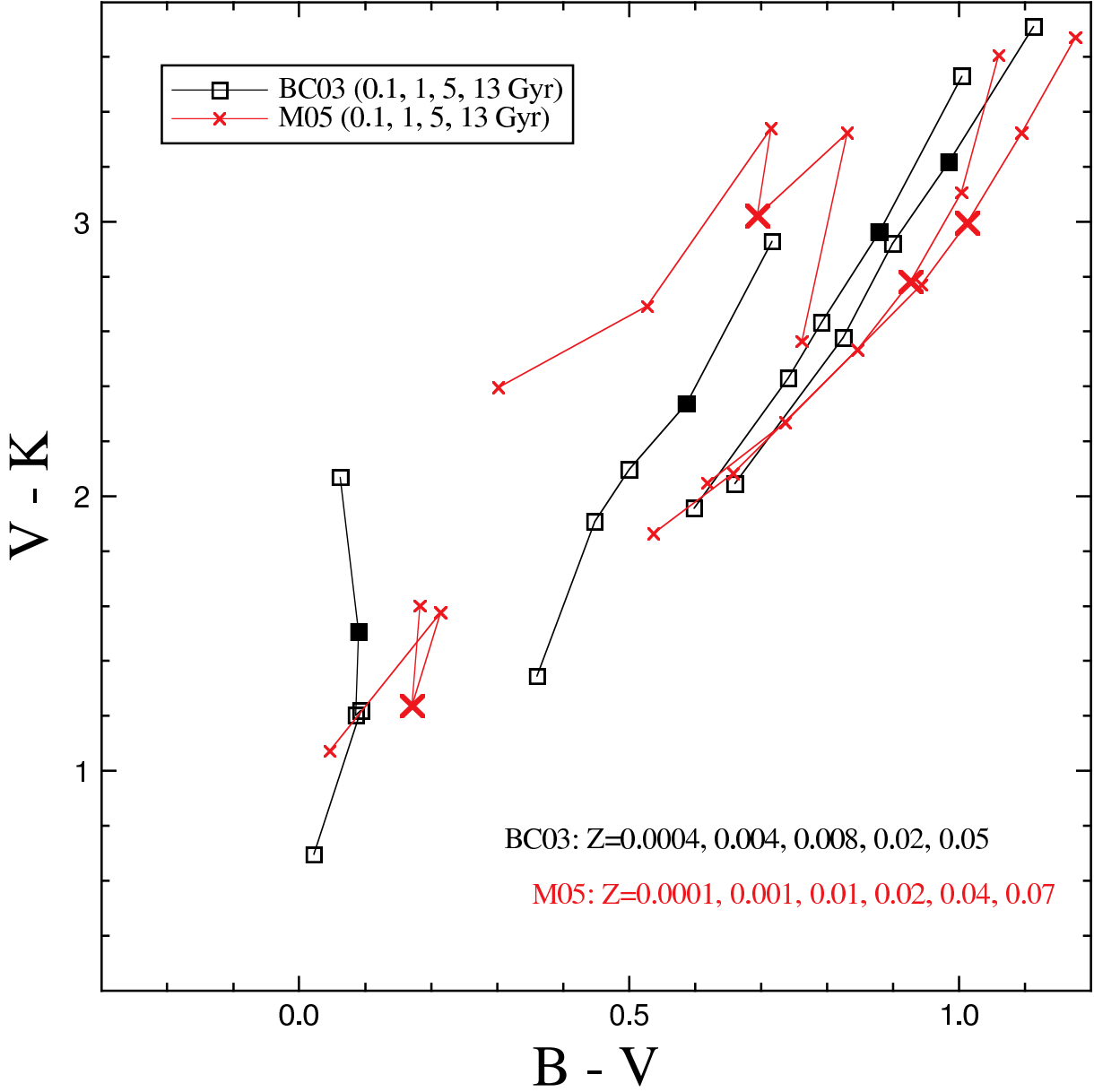}
\caption{Similar to Figure 6, but the Maraston (2005, M05) models 
(solid lines with crosses) are compared with BC03.
BC03's and M05's metallicities are indicated at the bottom 
right. At 100 Myr, M05 are only given at $Z = 0.001$, 0.01, 0.02, 
and 0.04.  To guide the eye, solar metallicity models are marked 
by filled or larger symbols.}
\end{figure}

\begin{figure}
\epsscale{.6}
\plotone{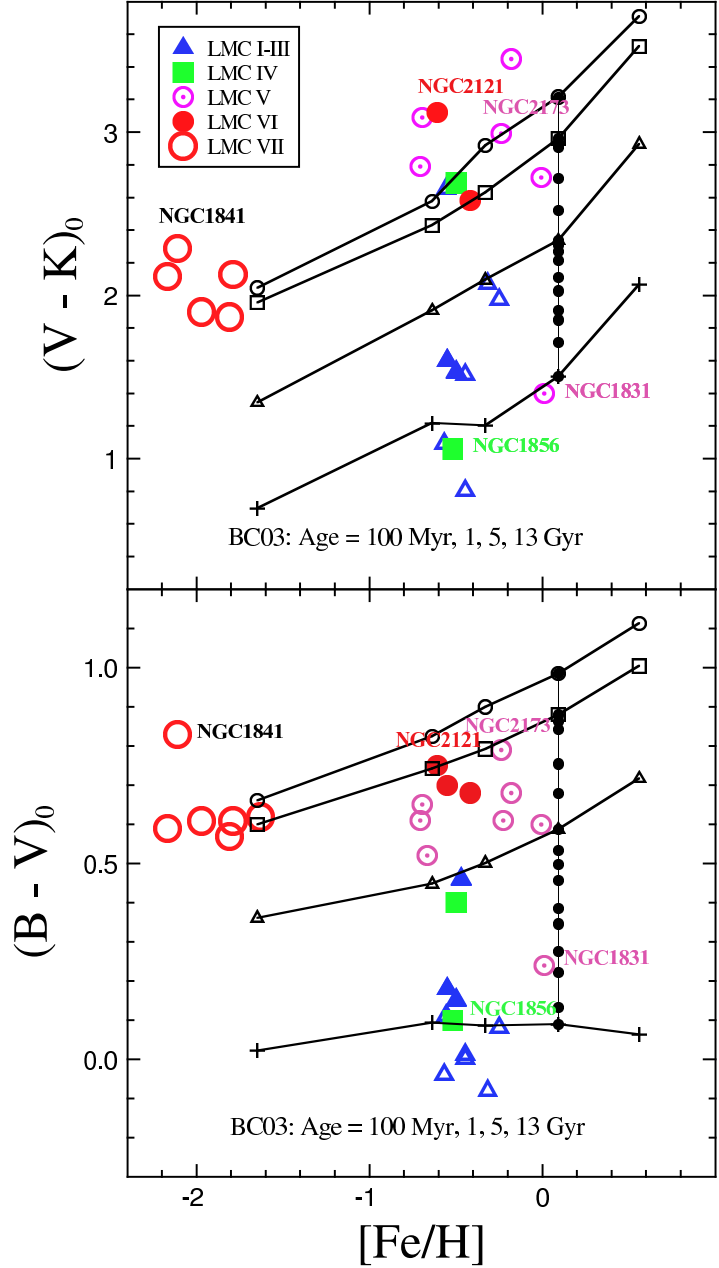}
\caption{LMC star clusters are compared with BC03 models at [Fe/H] 
vs. $(B - V)_{0}$ (bottom) and $(V - K)_{0}$ (top).  Among SWB type 
I-III, star clusters younger than 100 Myr are depicted with open triangles.  
At solar metallicity, eight 100 Myr increments between 100 Myr and 
1 Gyr, and 1.5, 2, 3, and 4 Gyr between 1 Gyr and 5 Gyr are connected with dots.
Both ($B - V$) and ($V - K$) colors of BC03 models are becoming 
{\em monotonically} redder with increasing age at given metallicity.  
Note that contrary to ($B - V$) colors, ($V - K$) colors show some evident 
mismatch between intermediate-age LMC star clusters and model predictions.  
Ages and metallicities of LMC clusters are taken from Pessev et al.\ (2006).  
Recent age estimates of NGC 1856 (300 Myr), NGC 1831 (700 Myr), 
NGC 2173 (1.6 Gyr), and NGC 2121 (2.9 Gyr) are from Kerber et al.\ (2007).  
NGC 1841 is indicated because of its unusually red color 
in $(B - V)_{0}$ (see text).}
\end{figure}

\begin{figure}
\epsscale{.6}
\plotone{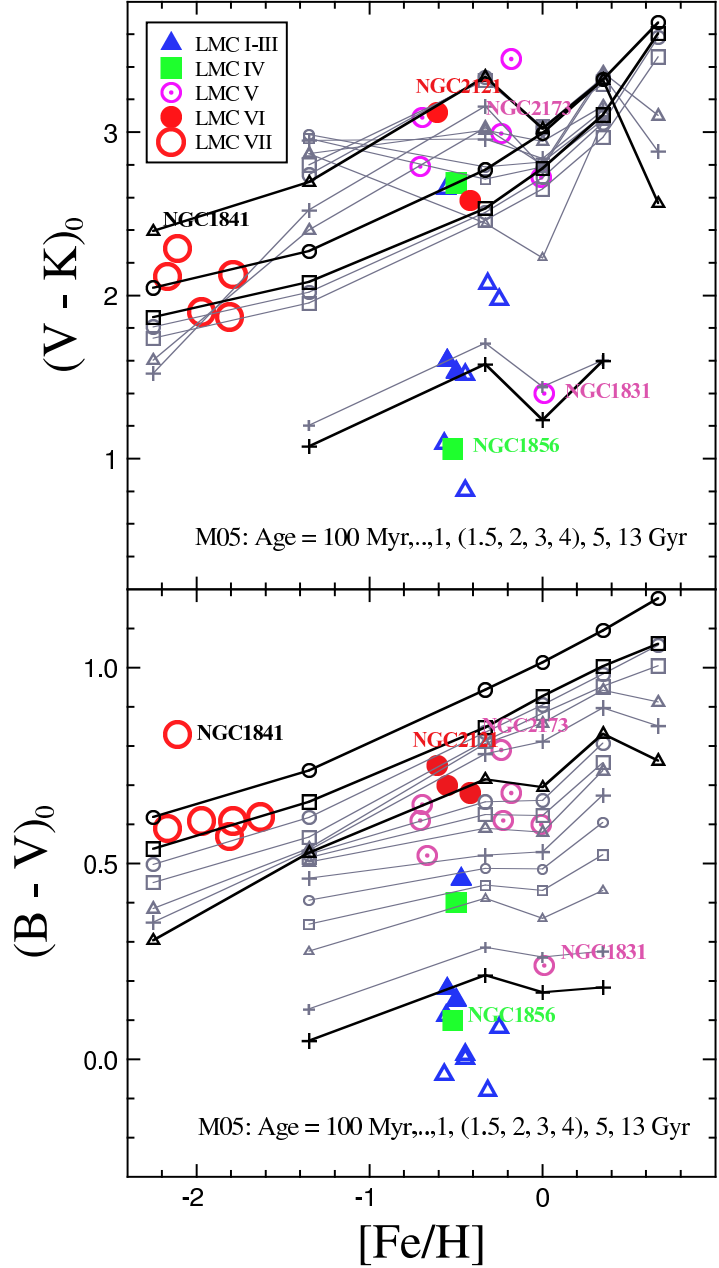}
\caption{Same as Figure 8, but M05 models are overlaid with LMC 
star clusters.  Four ages (100 Myr, 1, 5, 13 Gyr) that are shown 
at the Figure 7 are delineated with thicker lines with crosses, triangles, 
squares, and circles, respectively.  Between 100 Myr and 1 Gyr, 
every 100 Myr increment is depicted with the following symbol order: 
cross, triangle, square, and circle.  The same symbols are used for age 
increments between 1 Gyr and 5 Gyr.  While her ($B - V$) colors become 
monotonically redder with increasing age at given metallicity, her 
($V - K$) colors show a sudden leap at 300 Myr because of the onset of 
the TP-AGB.  M05 tuned her fuel consumption theorem parameters using the 
older age estimation of intermediate-age LMC star clusters.  
Recent age estimates of NGC 1856 (300 Myr), NGC 1831 (700 Myr), 
NGC 2173 (1.6 Gyr), and NGC 2121 (2.9 Gyr) are from Kerber et al.\ (2007).}
\end{figure}

\begin{figure}
\epsscale{.6}
\plotone{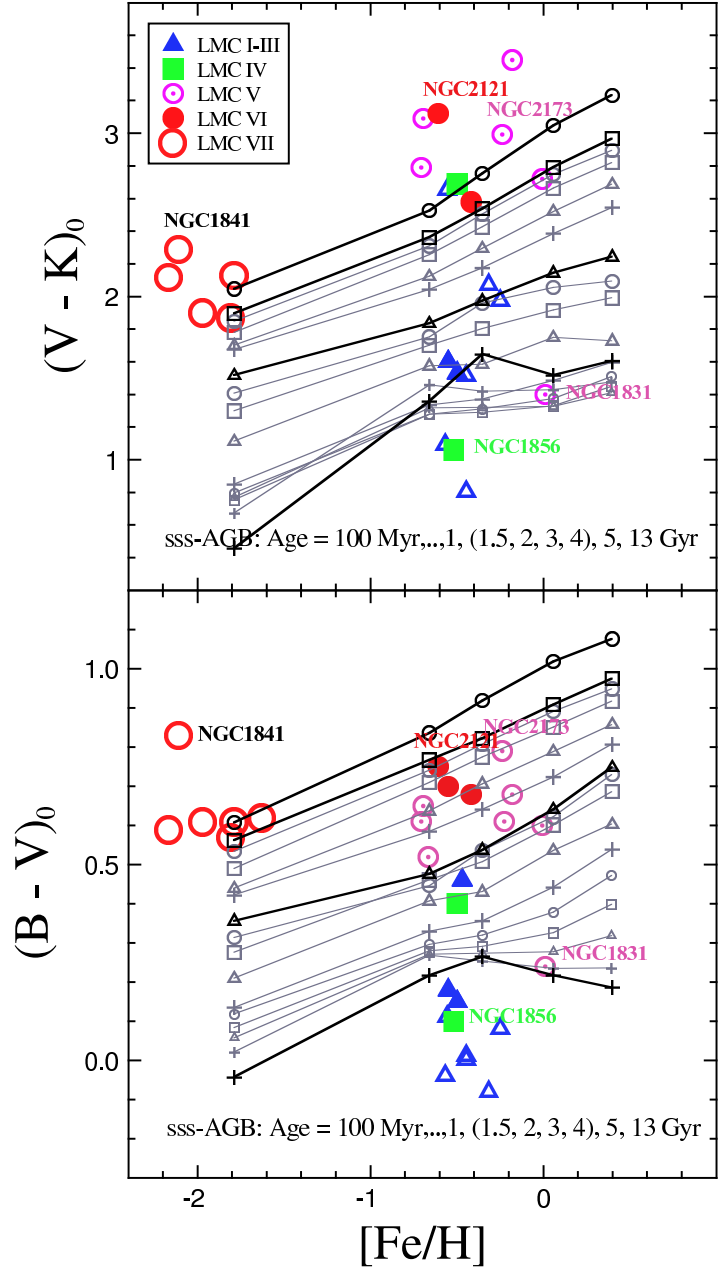}
\caption{Same as Figure 9, but our calculations with BaSTI-sss 
isochrones (without TP-AGB phases) are compared with LMC star clusters.  
Recent age estimates of NGC 1856 (300 Myr), NGC 1831 (700 Myr), 
NGC 2173 (1.6 Gyr), and NGC 2121 (2.9 Gyr) are from Kerber et al.\ (2007).  
Note that, contrary to ($B - V$) colors, ($V - K$) colors show some apparent 
mismatch between intermediate-age LMC star clusters and model predictions.}
\end{figure}

\begin{figure}
\epsscale{.6}
\plotone{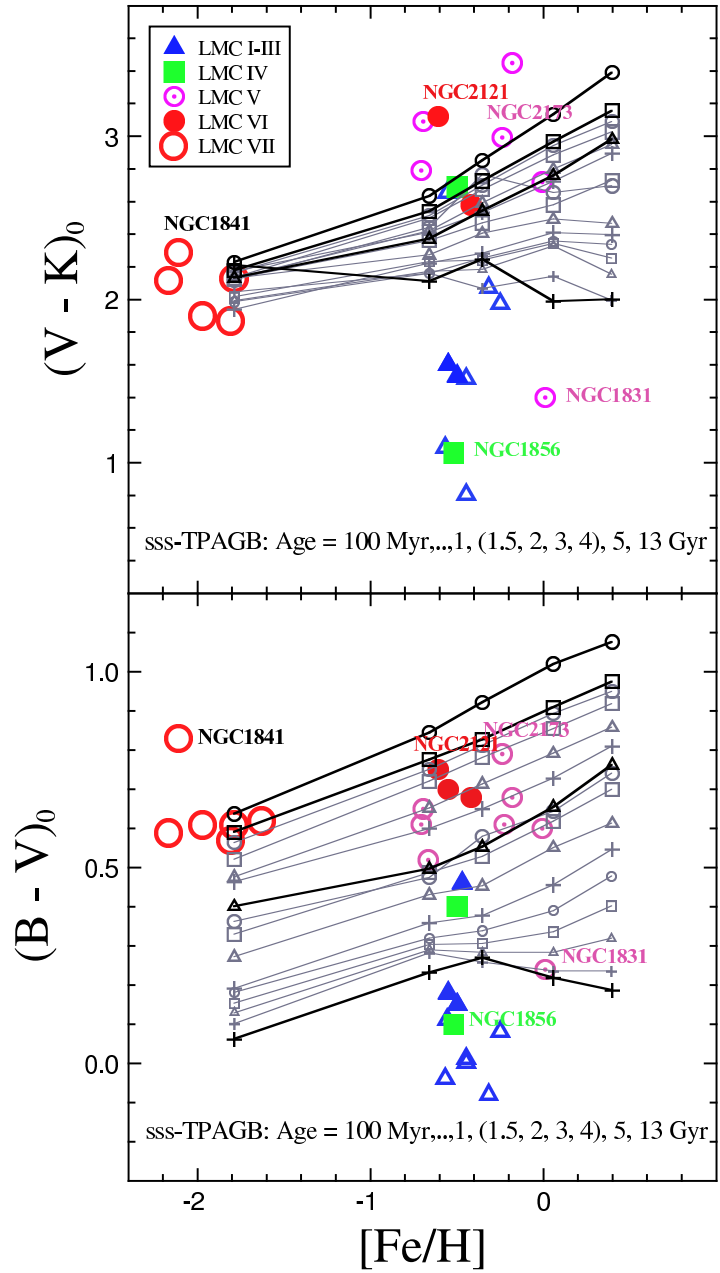}
\caption{Same as Figure 10, but our calculations with BaSTI-sss 
isochrones (with TP-AGB phases) are compared with LMC star clusters.  
Recent age estimates of NGC 1856 (300 Myr), NGC 1831 (700 Myr), 
NGC 2173 (1.6 Gyr), and NGC 2121 (2.9 Gyr) are from Kerber et al.\ (2007).  
Note that the ($V - K$) colors become significantly redder compared to 
Figure 10 at younger ages because of the TP-AGB phases although 
the mismatch between observations and models still lingers.}
\end{figure}

\begin{figure}
\epsscale{.6}
\plotone{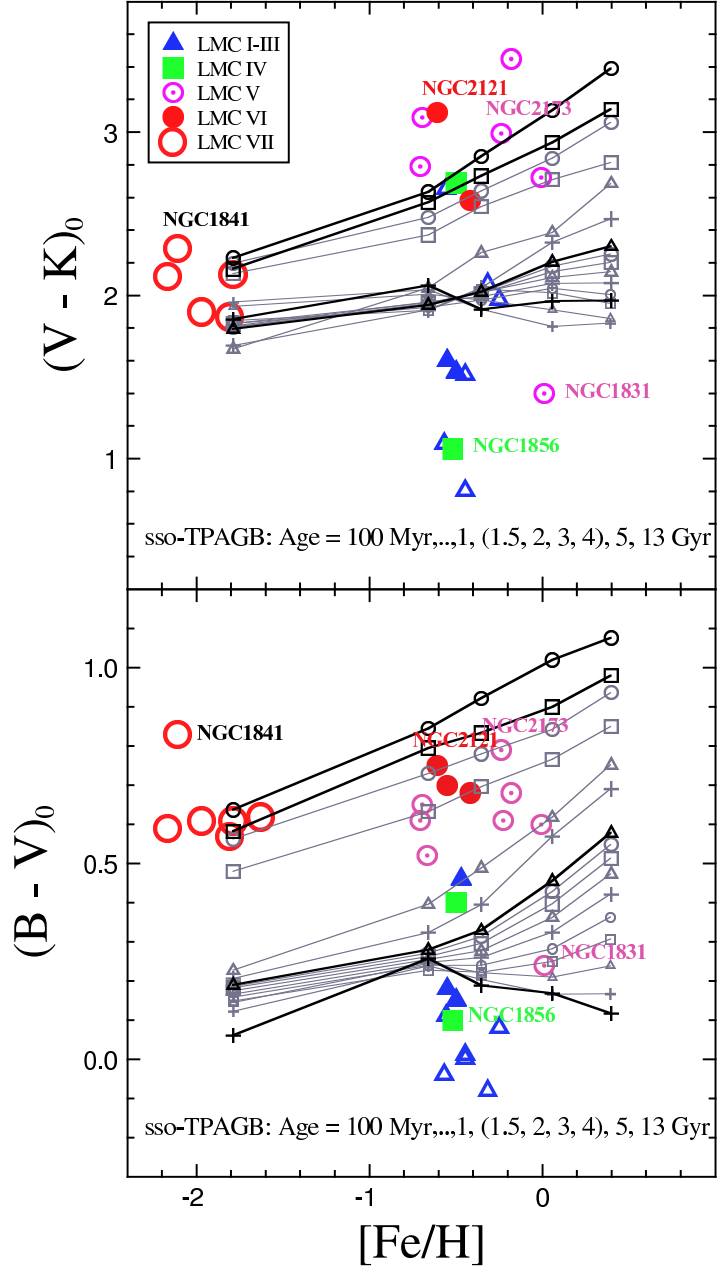}
\caption{Same as Figure 11, but our calculations with BaSTI-sso 
isochrones (with TP-AGB phases) are compared with LMC star clusters.
Note that both ($B - V$) and ($V - K$) colors become bluer compared to 
Figure 11 at younger ages because of the convective core overshooting effects.  
Recent age estimates of NGC 1856 (300 Myr), NGC 1831 (700 Myr), 
NGC 2173 (1.6 Gyr), and NGC 2121 (2.9 Gyr) are from Kerber et al.\ (2007).  
Although this case should be the most realistic one, it does not ameliorate 
the mismatch between observations and models in ($V - K$).}
\end{figure}

\begin{figure}
\epsscale{1.}
\plotone{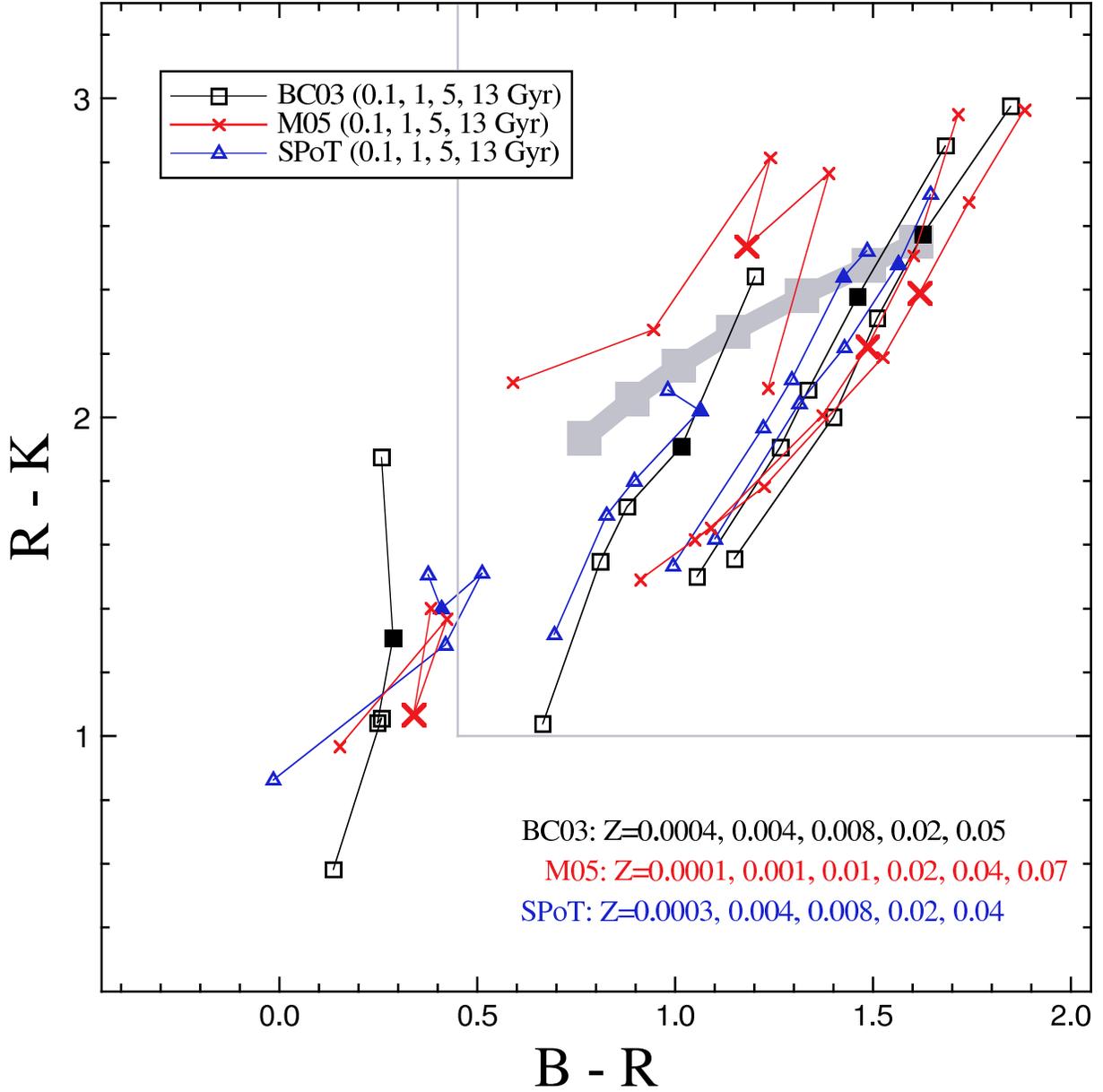}
\caption{Similar to Figures 6 and 7, but BC03, M05, and the SPoT 
simple stellar population models at 0.1, 1, 5, and 13 Gyr are 
compared in ($B - R$) vs.\ ($R - K$).  BC03's solar 
metallicity {\em composite} models from Figure 14 are shown here 
as the thick line with big filled squares (the seven average 
ages, $\langle A\rangle$) to compare the SSP and the CSP models.  
For the same purpose, the matching scales of Fig.\ 14 are delineated 
with gray straight lines.  According to Figure 11 of \citet{mac04}, 
spiral galaxies with {\em high} rotational velocity mostly swarm around 
this thick line.}
\end{figure}

\begin{figure}
\epsscale{.6}
\plotone{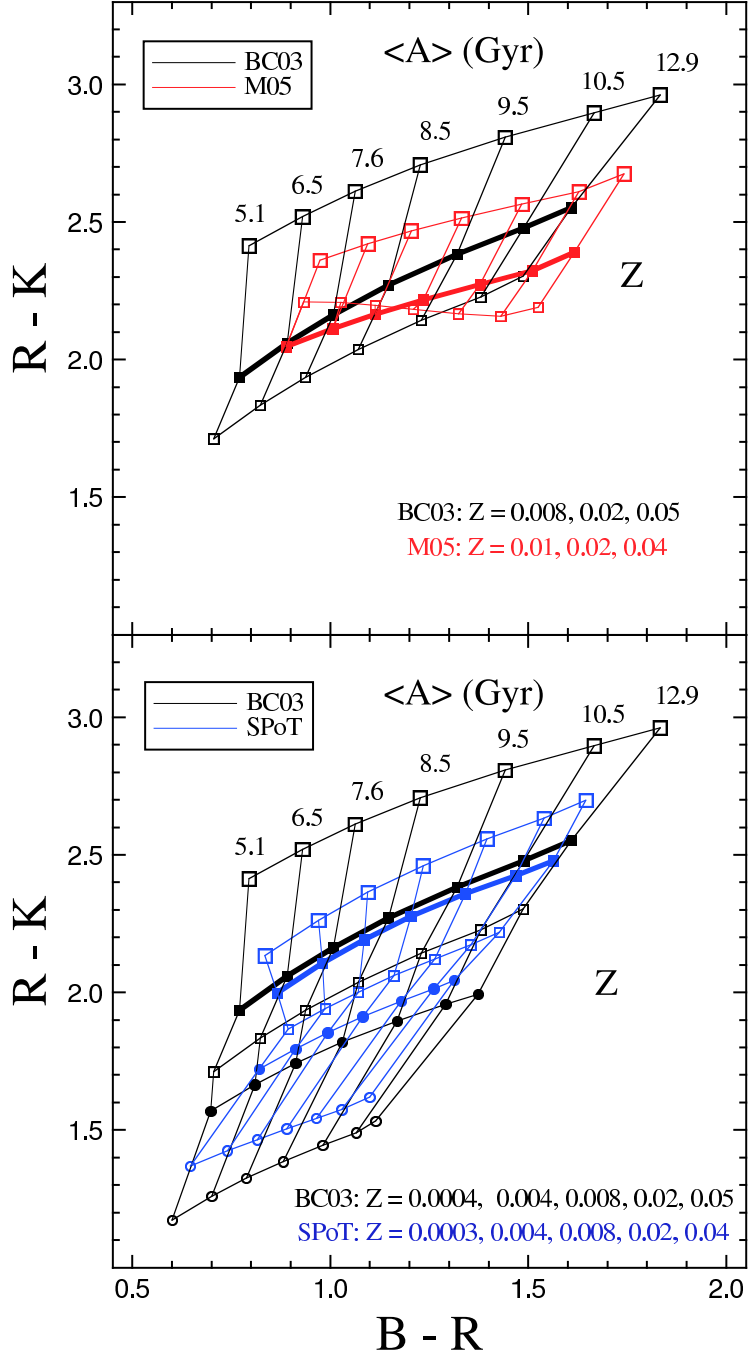}
\caption{Three different groups' {\em composite} stellar population 
models are generated as identical SFHs are employed.  
BC03 (black) and M05 (red) are compared in the top panel and 
BC03 and SPoT (blue) are contrasted in the bottom panel.
Same symbols are used for similar metallicities.  To guide the eye, 
solar metallicities are depicted with thicker lines.  Note also the 
compression of scale compared to the Figure 13 SSP illustration.}
\end{figure}

\begin{figure}
\epsscale{1.}
\plotone{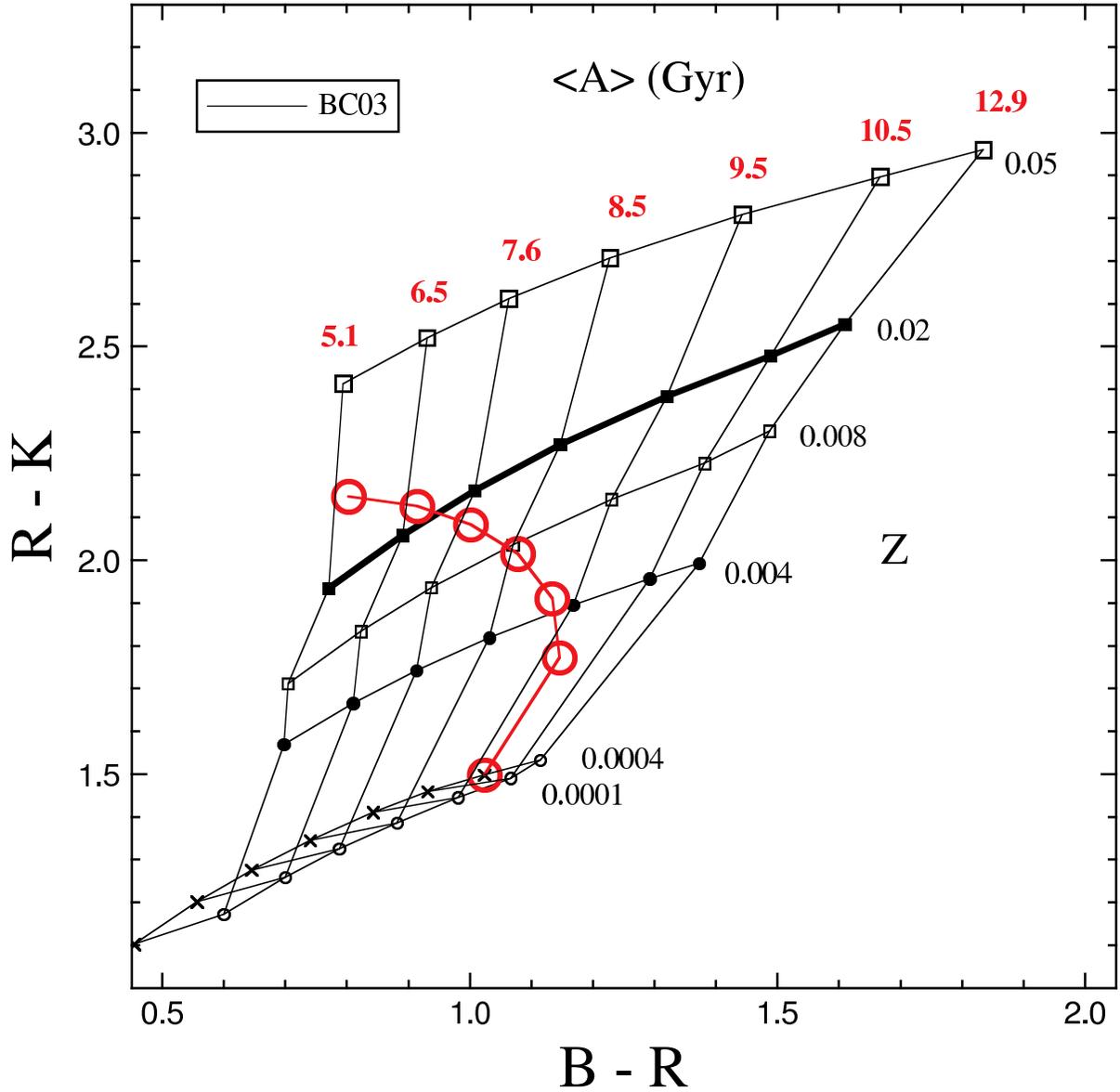}
\caption{Similar to Figure 14, with the BC03 models repeated (black), 
but {\em a chemical enrichment scheme with a monotonic age-metallicity 
relation} is incorporated in the BC03 composite stellar population models 
at given average ages (red circles; see text).}
\end{figure}

\end{document}